\documentclass[10pt,  twocolumn, twoside, letterpaper, conference, final]{IEEEtran}
\IEEEoverridecommandlockouts
\usepackage{cite}
\usepackage{amsmath,amssymb,amsfonts}
\usepackage{breq n}
\usepackage{algorithmic}
\usepackage{graphicx}
\usepackage{textcomp}
\usepackage{xcolor}
\usepackage{verbatim} 
\usepackage{balance}

\newtheorem{lemma}{\textbf{Lemma}}
\def\BibTeX{{\rm B\kern-.05em{\sc i\kern-.025em b}\kern-.08em
    T\kern-.1667em\lower.7ex\hbox{E}\kern-.125emX}}
    
\begin{document}
\title{Performance Analysis of NOMA Uplink Networks under Statistical QoS Delay Constraints}
\author{\IEEEauthorblockN{Mouktar Bello\IEEEauthorrefmark{1} $,$ Wenjuan Yu \IEEEauthorrefmark{2}, Arsenia Chorti\IEEEauthorrefmark{1} and Leila Musavian\IEEEauthorrefmark{4}} \\
\IEEEauthorblockA{\IEEEauthorrefmark{1}ETIS UMR8051, CY Université, ENSEA, CNRS, F-95000, Cergy, France}
\IEEEauthorblockA{\IEEEauthorrefmark{2}5GIC, Institute of Communication Systems, University of Surrey, Guildford, GU2 7XH, UK}

%\IEEEauthorblockA{\IEEEauthorrefmark{1} ETIS, Univ. Paris Seine, Univ. Cergy-Pontoise, ENSEA, CNRS, 95000 Cergy, FR}
\IEEEauthorblockA{\IEEEauthorrefmark{4}School of Computer Science and Electronic Engineering, University of Essex, Colchester, CO4 3SQ, UK}
}

%%SECond form Authors with Blocks
%\author{\IEEEauthorblockN{1\textsuperscript{st} Bello Mouktar}
%\IEEEauthorblockA{\textit{ETIS, Univ. Paris Seine, Univ. Cergy-Pontoise, \\ENSEA, CNRS, %95000 Cergy, FR}}
%\and
%\IEEEauthorblockN{2\textsuperscript{nd} Wenjuan Yu}
%\IEEEauthorblockA{\textit{5GIC, Institute of Communication Systems,\\ University of Surrey, %Guildford, GU2 7XH, UK}}

%\and
%\IEEEauthorblockN{3\textsuperscript{rd} Arsenia Chorti}
%\IEEEauthorblockA{\textit{ETIS, Univ. Paris Seine, Univ. Cergy-Pontoise, \\ENSEA, CNRS, %95000 Cergy, FR}}

%\and
%\IEEEauthorblockN{4\textsuperscript{th} Leila Musavian}
%\IEEEauthorblockA{\textit{School of Computer Science and Electronic Engineering,\\ %University of Essex, Colchester, CO4 3SQ, UK}}
%}

\maketitle

\begin{abstract}
In the fifth generation and beyond (B5G), delay constraints emerge as a topic of particular interest, e.g. for ultra-reliable low latency communications (URLLC) such as autonomous vehicles and enhanced reality. In this paper, we study the performance of a two-user uplink NOMA network under statistical quality of service (QoS) delay constraints, captured through each user's effective capacity (EC). We propose novel closed-form expressions for the EC of the NOMA users and show that in the high signal to noise ratio (SNR) region, the ``strong'' NOMA user has a limited EC, assuming the same delay constraint as the ``weak'' user. We demonstrate that for the weak user, OMA achieves higher EC than NOMA at small values of the transmit SNR, while NOMA outperforms OMA in terms of EC at high SNRs. On the other hand, for the strong user the opposite is true, i.e., NOMA achieves higher EC than OMA at small SNRs, while OMA becomes more beneficial at high SNRs. This result raises the question of introducing ``adaptive" OMA / NOMA policies, based jointly on the users' delay constraints as well as on the available transmit power. 
\end{abstract}

\begin{IEEEkeywords}
NOMA, QoS, low latency, effective capacity, B5G.
\end{IEEEkeywords}

\section{Introduction}
\textcolor{black}{Non-orthogonal} multiple access (NOMA) schemes have attracted a lot of attention recently, allowing multiple  users to be served simultaneously with enhanced spectral efficiency; it is known that the boundary of achievable rate pairs (in the case of two users) using NOMA is outside the capacity region achievable with orthogonal multiple access (OMA) techniques \cite{islam2016power} or other schemes \cite{Chorti_FTN}. Superior achievable rates are attainable \textcolor{black}{through} the use of superposition coding at the transmitter and of successive interference cancellation  (SIC) at the receiver \cite{saito2013non,yu2018link}. The SIC receiver decodes multi-user signals with descending received signal power and subtracts the decoded signal(s) from the received \textcolor{black}{superimposed} signal, so as to improve the signal-to-interference ratio. The process is repeated until the signal of interest is decoded. In uplink NOMA networks, the strongest user's signal is decoded first (as opposed to downlink NOMA networks in which the inverse order is applied).  

Besides, in a number of emerging applications, \textcolor{black}{delay quality of service (QoS) becomes} increasingly important, e.g., ultra reliable low latency communication (URLLC) systems.  Furthermore, in future wireless networks, users are expected to necessitate flexible delay guarantees for achieving different service requirements. In order to satisfy diverse delay requirements, a simple and flexible delay QoS model is imperative to be applied and investigated. In this respect, the effective capacity (EC) theory can be employed \cite{yu2016tradeoff},\cite{wu2003effective} \cite{tang2007cross}, with EC denoting the maximum constant arrival rate which can be served by a given service process, while guaranteeing the required statistical delay provisioning. We studied the delay-constrained communications for a downlink NOMA network in \cite{yu2018link} and with secrecy constraints \cite{Chorti_asilomar} in \cite{Wenjuan_Ersi}. The present analysis complements \cite{yu2018link}, focusing on uplink transmissions. NOMA, as a more spectrum-efficient technique, is considered to be promising for supporting the massive number of devices to access the uplink connections. Hence, we believe that it is important to investigate the delay-constrained achievable rate for an uplink NOMA network. 

In this paper, we provide a performance evaluation of the uplink \textcolor{black}{transmission for} a two-user NOMA network under delay constraints, captured through the users' \textcolor{black}{ECs}. We note that the EC is a QoS aware \textcolor{black}{data-link} layer metric \cite{wu2003effective}, that captures the achievable rate under a delay violation probability threshold.  

In this work, we first derive novel closed-form expressions for the ECs of both users; we then provide four Lemmas for the asymptotic performance of the network with NOMA and OMA. The conclusions drawn are supported by an extensive set of simulations. The paper is organized as follows. In Section II we investigate the EC of a two user uplink NOMA  system under the delay QoS constraints. Simulation results are given in Section III, followed by conclusions in Section IV.

\section{Effective Capacity of Two-user NOMA Uplink Network}
Assume a two-user NOMA uplink network with users U$_1$ and U$_2$ in a Rayleigh block fading propagation channel, with respective channel gains during a transmission block denoted by  $|h_1|^2<|h_2|^2$. The users transmit corresponding symbols $S_1, S_2$ respectively, with power
$\mathbb{E}[|S_i|^2]=P_i, i=1,2$ and the total power $P_T=\sum_{i=1}^{2}{P_i}=1$. Here, $P_i$ is the power coefficient for the user $i$ and normalized transmission powers are assumed\cite{yang2016uplinknoma}. The received superimposed signal can be expressed as\cite{nzhang2016uplink}
\begin{equation}
    Z = \sum_{i=1}^{2}\sqrt{P_i}h_iS_i+w,
\end{equation}
where \textcolor{black}{$w$} denotes a zero mean circularly symmetric complex Gaussian random variable with variance $\sigma^2$.  
The receiver will first decode the symbol of the strong user treating the transmission of the weak as interference. After decoding it, \textcolor{black}{the receiver will} suppress it from $Z$ and decode the signal of the weak user. Following the SIC principle and denoting by \(\rho = \frac{1}{\sigma^2}\) the transmit SNR, the achievable rates, in b/s/Hz, for user
U$_i, i=1,2$, is expressed as: \cite{fan2015uplink}
 \begin{equation}
    R_i= \log_2 \left[1+ \frac{\rho P_i|h_i|^2}{1+\rho \sum_{l=1}^{i-1}P_l|h_l|^2}\right].
 \end{equation}
 
To clarify further, let $\theta_i$ \textcolor{black}{be} the statistical delay QoS exponent of the $i$-th user, and assume that the service process satisfies the G\"{a}rtner-Ellis theorem \cite{wu2003effective}. The delay exponent $\theta_i$ captures how strict the delay constraint is \cite{wu2003effective}. A slower decay rate can be represented by a smaller $\theta_i$, which indicates that the system is more delay tolerant, while a larger $\theta_i$ corresponds to a system with more stringent QoS requirements. 
Applying the EC theory in a uplink NOMA with two users, the $i$-th user's EC over a block-fading channel, is defined as:
\begin{align}
      E_c^i= -\frac{1}{\theta_i T_{{f}} B} \ln \left( \mathbb{E}\left[e^{-\theta_i T_{\text{f}} B R_i}\right] \right) \quad \left(\text{in b/s/Hz}\right),
      \label{eq:ECdefinition}
\end{align}
where $T_f$ is the fading-block length, $B$ is the bandwidth and $\mathbb{E}\left[\cdot\right]$ denotes expectation over the channel gains. 
By inserting $R_i$ into (\ref{eq:ECdefinition}), we obtain the following expression for the EC of the $i$-{th} user
\begin{align}
E_c^i &=& \frac{1}{\beta_i}\log_2\left(\mathbb{E}\left[(1+ \frac{\rho P_i|h_i|^2}{1+\rho \sum_{l=1}^{i-1}P_l|h_l|^2})^{\beta_i}\right]\right)
\end{align}
where $\beta_i=-\frac{\theta_iT_fB}{\ln{2}}$, $ i=1,2$, is the normalized (negative) QoS exponent.

\subsection{ECs in a Two-user NOMA Uplink Network}
For the ordering of the channel gains we make use of the theory of order statistics in the following analysis \cite{yang2011order}.

Assuming a Rayleigh wireless environment, the channel gains, denoted by $x_i=|h_i|^2, i=1,2$, are exponentially distributed with probability density function (PDF) and cumulative density function (CDF) respectively given by $f(x_i)=e^{-x_i}$, $F(x_i)=1-e^{-x_i}$. 

Then, according to order statistics\cite{yang2011order}, the ordered channel gains have respective PDFs $f_{i:2}(x_i), i=1,2$, and joint PDF $f(x_1, x_2)$ that are expressed as
\begin{align}
f_{1:2}(x_1)&=2e^{-2 x_1},\\
f_{2:2}(x_2)&=2e^{-x_2}\left(1-e^{-x_2}\right),\\
   f(x_1, x_2) &= 2 e^{- x_{1}}e^{- x_{2}}.
   \label{eq:joint pdf 2 user}
  \end{align}

As a result, the EC of User 1, denoted by $E_c^1$ is expressed as
\begin{eqnarray}
    E_c^1 &=&\frac{1}{\beta_1}\log_2(\mathbb{E}[(1+ \rho P_1 x_{1})^{\beta_1}])\nonumber
    \\ 
    &=& \frac{1}{\beta_1}\log_2\Big(\int_{0}^{\infty}(1+ \rho P_1 x_1)^{\beta_1}f_{{1:2}}(x_1) dx_1\Big) \nonumber\\
    &=&\frac{1}{\beta_1}\log_2\Big(\frac{2}{P_1 \rho}\times U\left(1,2+\beta_1,\frac{2}{\rho P_1}\right)\Big).
    \label{eq:EC1}
\end{eqnarray}
where $U(\cdot, \cdot, \cdot)$ denotes the confluent hypergeometric function \cite{yu2018link}.
On the other hand, the EC of the User 2 is evaluated as 
\begin{align}
  E_c^2 &= \frac{1}{\beta_2}\log_2\left(\mathbb{E}\left[\left(1+ \frac{\rho P_2 x_{2}}{1+ \rho P_1 x_{1}}\right)^{\beta_2}\right]\right)\nonumber \\
  &=\frac{1}{\beta_2}\log_2\left(\int_{0}^{\infty}\!\!\!\!\int_{x_{1}}^{\infty}\!\!\!\!\Big(1+ \frac{\rho P_2 x_2}{1+\rho P_1 x_1}\Big)^{\beta_2}f(x_1,x_2) dx_2 dx_1\right) \nonumber\\
  &=\frac{1}{\beta_2}\log_2\Big(2 P_2^{1-\beta_2}(\rho P_2)^{\beta_2} e^{\frac{1}{\rho P_2}}e^{-\frac{(P_1-P_2)}{\rho P_2}}\Big) \nonumber \\
  &+\frac{1}{\beta_2}\log_2\Bigg(\sum_{j=0}^{-\beta_2}\binom{-\beta_2}{j}(\rho P_1)^{j}\times \sum_{k=0}^{\infty}\frac{(-1)^k (P_2-P_1)^k}{k!(1+j+k)}\nonumber \\
  &\times\Big[\Gamma[2+\beta_2+j+k,\frac{1}{\rho P_2}]\nonumber \\
  &-(\rho P_2)^{-1-j-k}\Gamma[1+\beta_2,\frac{1}{\rho P_2}] \Big] \Bigg),
  \label{eq:EC2}
\end{align}
with $\Gamma(\cdot, \cdot)$ denoting the incomplete Gamma function \cite{yu2018link}.

The proof for deriving $E_c^1$ is omitted due to space limitations while
for $E_c^2$ is provided in Appendix I.%\footnote{All proofs are in the arxiv}. 
%\newline

%\subsection{EC of a Two-Users OMA Network}
In order to perform a comparative performance analysis, here we provide the achievable data rates for a two-user OMA network, denoted by $\widetilde{R}_{i}, i=1,2$, given as
\begin{equation}
  \widetilde{R}_{i}=\frac{1}{2}\log_2\Big(1+\rho P_T|h_i|^2\Big), i=1,2
\end{equation}
Note that $\frac{1}{2}$ is due to the equal allocation of resources to both users. The corresponding expressions are obtained for the ECs of both users in a OMA network, denoted by $\widetilde{E}_c^{i}$, given as

\begin{align}
\widetilde{E}_c^{i}&=\frac{1}{\beta_i}\log_2\Big(\mathbb{E}\Big[(1+\rho P_T  |h_i|^2)^{\frac{\beta_i}{2}}\Big] \Big)\label{eq:EC1OMA} \nonumber\\
i.e,\\\nonumber
&\widetilde{E}_c^{1}=\frac{1}{\beta_1}\log_2\left(\frac{2}{\rho }\times U\left(1, 2+\frac{\beta_1}{2}, \frac{2}{\rho}  \right)\right)\\\nonumber
&\!\!\!\!\!\widetilde{E}_c^{2}=\frac{1}{\beta_2}\log_2\left(\frac{2}{\rho } \sum_{k=0}^{1}\binom{1}{k}(-1)^k \times U\left(1, 2+\frac{\beta_2}{2}, \frac{1+k}{\rho}  \right)\right)
\end{align}
The proof is omitted due to space limitations.

\subsection{Asymptotic Analysis}
We first perform an asymptotic analysis with respect to the SNR. Our results are summarized in Lemma 1. 

\begin{lemma} \textit{In the low and high SNR regimes, respectively, the following conclusions hold:
        \begin{enumerate}
            \item When $\rho\rightarrow0$, then, $E_c^1\rightarrow0$, $E_c^2\rightarrow0$, $\widetilde{E}_c^{1}\rightarrow0$, $\widetilde{E}_c^{2}\rightarrow0$, $E_c^1-\widetilde{E}_c^{1}\rightarrow0$, $E_c^2-\widetilde{E}_c^{2}\rightarrow0$;
            \item When $\rho\rightarrow +\infty$, then $E_c^1\rightarrow+\infty$, $E_c^2\rightarrow \frac{1}{\beta_2}\log_2\left(\mathbb{E}\left[\left(1+\frac{P_2 |h_2|^2}{P_1 |h_1|^2}\right)^{\beta_2}\right]\right)$, $\widetilde{E}_c^{1}\rightarrow+\infty$, $\widetilde{E}_c^{2}\rightarrow+\infty$, $E_c^1-\widetilde{E}_c^{1}\rightarrow+\infty$, $E_c^2-\widetilde{E}_c^{2}\rightarrow-\infty$.
        \end{enumerate}
}
\end{lemma}
\begin{IEEEproof}: The proof is provided in Appendix II.\end{IEEEproof}

Lemma 1 indicates that the ECs of both users are vanishingly small at low values of  $\rho$, irrespective of employing NOMA or OMA. On the other hand, at high SNRs, we notice that the EC of the strong user with NOMA is limited to a finite value. On the contrary, for the weaker user, when $\rho >> 1$, its achievable EC in the NOMA uplink increases without bound. This is the exact opposite of the downlink scenario, where it is the weaker user which is limited in terms of EC, when $\rho >> 1$ \cite{yu2018link}.

Now, the question is how the ECs evolve with $\rho$ between the two asymptotic regimes. To answer this question and to further analyze the impact of $\rho$ on the individual EC, we look at the derivatives with the respect of $\rho$ \cite{yu2018link}.
\begin{lemma} \textit{For the EC of the User 1, in a two-user uplink network the following hold: 
             \begin{enumerate}
              \item $ \frac{\partial E_c^1}{\partial\rho}\ge0$ and $\frac{\partial\widetilde{E}_c^{1}}{\partial\rho}\ge0$, $\forall \rho$;
              \item When $\rho\rightarrow0$, then $\lim\limits_{\rho\rightarrow0}(\frac{\partial (E_c^1-\widetilde{E}_c^{1})}{\partial\rho})=\frac{P_1-\frac{1}{2}}{\ln 2}\mathbb{E}[|h_1|^2]$;
              \item When $\rho>> 1$, then $\frac{\partial (E_c^1-\widetilde{E}_c^{1})}{\partial\rho}\approx\frac{1}{2\rho \ln2}\ge0$ and it approaches $0$ when $\rho\rightarrow\infty$.
             \end{enumerate}
}
        \end{lemma}
       \begin{IEEEproof}: The proof is provided in Appendix III.\end{IEEEproof}

    Lemma 2 indicates that for User 1, when the transmit SNR $\rho$ is very small, the EC with OMA increases faster than the EC with NOMA. On the other hand, Lemma 2 shows that when the transmit SNR is very large, the EC with NOMA increases faster than with OMA. 
    
    Combining Lemma 2 and Lemma 1, we can conclude that, $E_c^1-\widetilde{E}_c^{1}$ starts at vanishingly small value, first decreases, and subsequently increases to $\infty$ at a gradually reducing speed. This means that for the weaker user, OMA achieves higher EC than NOMA at small values of the transmit SNR $\rho$. At high values of $\rho$, NOMA becomes more beneficial for the weak user. Finally, when $\rho\rightarrow\infty$ the performance gain of NOMA over OMA reaches a constant value in the case of User 1.

\begin{lemma} \textit{For the EC of the User 2, in a two-user uplink network the following hold: 
             \begin{enumerate}
              \item $ \frac{\partial E_c^2}{\partial\rho}\ge0$ and $ \frac{\partial \widetilde{E}_c^{2}}{\partial\rho}\ge0$, $\forall \rho$;
              \item When $\rho\rightarrow0$, then $\lim\limits_{\rho\rightarrow0}(\frac{\partial (E_c^2-\widetilde{E}_c^{2})}{\partial\rho})=\frac{P_2}{2\ln 2}\mathbb{E}[|h_2|^2]$
              \item When $\rho>> 1$, then $\frac{\partial (E_c^2-\widetilde{E}_c^{2})}{\partial\rho}\approx-\frac{1}{2 \ln2}\frac{1}{\rho}<0$ and it approaches 0 when $\rho\rightarrow\infty$.
             \end{enumerate}
      }
        \end{lemma}
      \begin{IEEEproof}  The proof is provided in Appendix IV.\end{IEEEproof}
      
Lemma 3 indicates that, for User 2, when the transmit SNR $\rho$ is very small, the uplink EC with NOMA increases faster than that with OMA. On the other hand, when the transmit SNR is very large, the uplink EC with OMA increases faster than that with NOMA. Combining Lemma 3 and Lemma 1, we can conclude that, $E_c^2-\widetilde{E}_c^{2}$ starts at an initial vanishingly small value, first increases, and subsequently decreases to $-\infty$ with a gradually diminishing rate. This means that for the stronger user, NOMA achieves higher EC than OMA at small values of the transmit SNR $\rho$. At high values of $\rho$, OMA becomes more beneficial for the strong user. Finally, when $\rho\rightarrow\infty$ the performance gain of OMA over NOMA reaches a constant value, for the stronger user.
    
    Finally, we investigate the sum ECs when using OMA and NOMA, denoted by $V_N$ and $V_O$,
    \begin{eqnarray}
        V_N&=&E_c^1+E_c^2,\\
        V_O&=&\widetilde{E}_c^1+\widetilde{E}_c^2.
    \end{eqnarray}
    Our conclusions are drawn in Lemma 4.
\begin{lemma} \textit{For the sum EC with NOMA, denoted by $V_N$, and with OMA, denoted by $V_O$, in a two-user uplink network, the following hold:
              \begin{enumerate}
              \item $ \frac{\partial V_N}{\partial\rho}\ge0$ and $\frac{\partial V_O}{\partial\rho}\ge0$, $\forall \rho$;
              \item When $\rho\rightarrow0$, $V_N\rightarrow0$, $ \lim\limits_{\rho\rightarrow 0}(\frac{\partial V_N}{\partial\rho})=\frac{P_1}{\ln2}\mathbb{E}[|h_1|^2]+\frac{P_2}{ \ln2}\mathbb{E}[|h_2|^2]\ge0$, and $V_O\rightarrow0$, $ \lim\limits_{\rho\rightarrow 0}(\frac{\partial V_O}{\partial\rho})=\frac{P_1}{2 \ln2}\mathbb{E}[|h_1|^2]+\frac{P_2}{ 2\ln2}\mathbb{E}[|h_2|^2]\ge0$;
              \item When $\rho>>1$, $V_N\rightarrow\infty$, $\lim\limits_{\rho\rightarrow\infty}(\frac{\partial V_N}{\partial\rho})=0$, and $V_O\rightarrow\infty$, $\lim\limits_{\rho\rightarrow\infty}(\frac{\partial V_O}{\partial\rho})=0$.
     %         \end{enumerate}
    %Considering the total EC in OMA, $V_O$, for the two user system, we prove that:
   % \begin{enumerate}
              %\item $\rho$, $ \frac{\partial V_O}{\partial\rho}\ge0$, 
              %\item When $\rho\rightarrow0$, 
              %\item When $\rho\rightarrow\infty$, $V_O\rightarrow\infty$ $\lim\limits_{\rho\rightarrow\infty}(\frac{\partial V_O}{\partial\rho})=0$
             \end{enumerate}
}
\end{lemma}
\begin{IEEEproof} The proof is provided in Appendix V.\end{IEEEproof}

Lemma 4 indicates that when NOMA is applied, the sum EC has a constant increasing rate at small value of the transmit SNR $\rho$ that depends on the average of the channel power gains and the allocated power coefficients. A similar conclusion is reached when using OMA. On the other hand, when $\rho>>1$, Lemma 4 indicates that the rate at which the sum ECs increase reaches a plateau, both in the case of NOMA and OMA.

\section{Numerical Results}

In this section, the Lemmas presented in Section II will be validated through Monte Carlo simulations.
We consider a two user uplink NOMA system, with the following settings: normalized transmission powers for both users, $P_1=0.2$, $P_2=0.8$, normalized delay exponent $\beta_1=\beta_2=-1$ for both users, unless otherwise stated.

\begin{figure}[t]
  \begin{center}
    \includegraphics[width=0.5\textwidth]{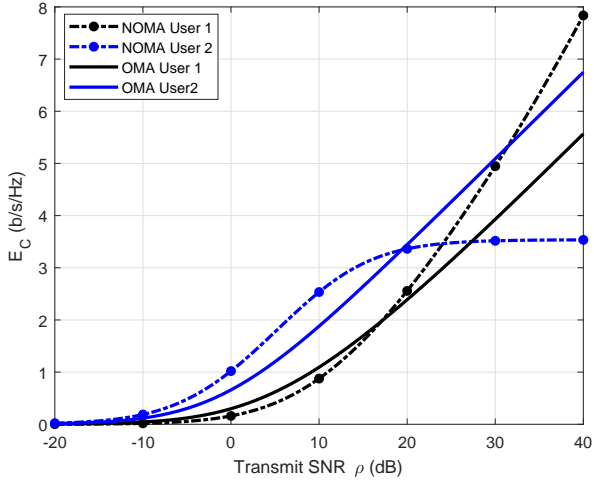}
\caption{
$E_c^1$, $E_c^2$in a two-user NOMA uplink network compared to Ecs of two users OMA, versus $\rho$}
      \label{fig6}
    \end{center}
\end{figure}

In Fig. \ref{fig6} the ECs of the  two-user uplink NOMA and OMA networks are depicted  versus the transmit SNR. We note that for the weak user, OMA is advantageous than NOMA for low transmit SNRs, and NOMA is advantageous OMA at high transmit SNRs. Reverse conclusions can be drawn for the strong user. We notice also that the EC of the strong user converges at high SNRs. This provides numerical validation for Lemma 1. 
 
\begin{figure}[t]
     \begin{center}
      \includegraphics[width=0.5\textwidth]{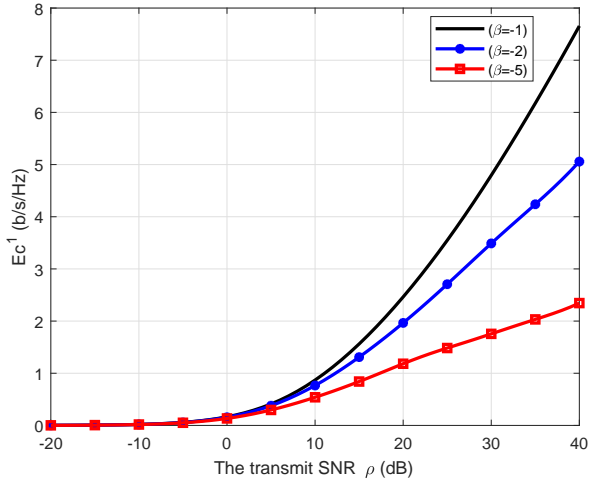}
      \caption{$E_c^1$ versus the transmit SNR, for several delays.}
      \label{fig7}
       \end{center}
  \end{figure}

\begin{figure}[t]
       \begin{center}
      \includegraphics[width=0.5\textwidth]{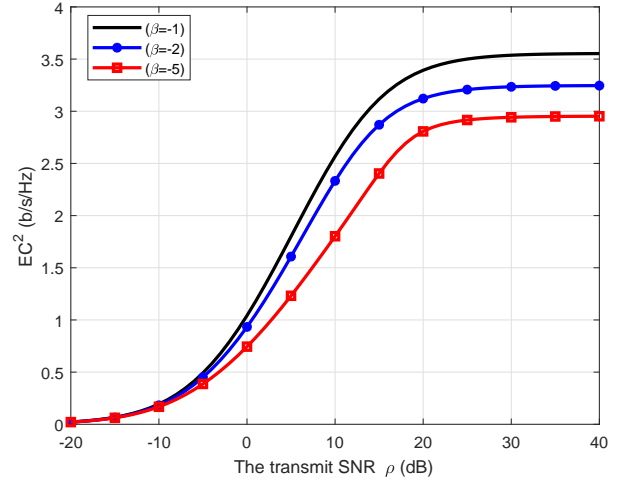}
      \caption{$E_c^2$ versus the transmit SNR $\rho$ for several delays.}
      \label{fig8}
       \end{center}
  \end{figure}

 Figs. \ref{fig7} and \ref{fig8}, show respectively the EC of User 1 and User 2, versus the transmit SNR, for different values of $\beta_1=\beta_2=\beta$. When the delay constraints become more stringent, i.e., $\beta$ decreases (equivalently, $\theta$ increases), the individual link-layer rates in NOMA decrease, for both users.

\begin{figure}[t]
     \begin{center}
      \includegraphics[width=0.5 \textwidth]{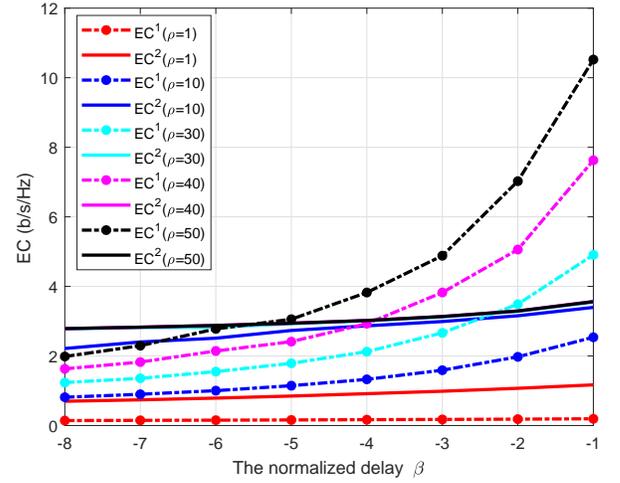}
      \caption{$E_c^1$ and $E_c^2$ in a two-user NOMA compared to ECs of two users OMA, versus normalized delay $\beta$, for different values of $\rho$.}
      \label{fig10}
     \end{center}
  \end{figure}

In Fig. \ref{fig10}, the ECs of the strong and weak users are depicted across different SNR values, ($\rho\in\{1, 10, 30, 40, 50\}$ dB, as functions of the (negative) normalized delay exponent, for NOMA and OMA scenarios. We notice that the EC of each user is identical for NOMA and OMA, for small and large values of the normalized delay exponent. And with the transmit SNR $\rho$ increasing, the EC increases for both users.

\begin{figure}[t]
       \begin{center}
      \includegraphics[width=0.5\textwidth]{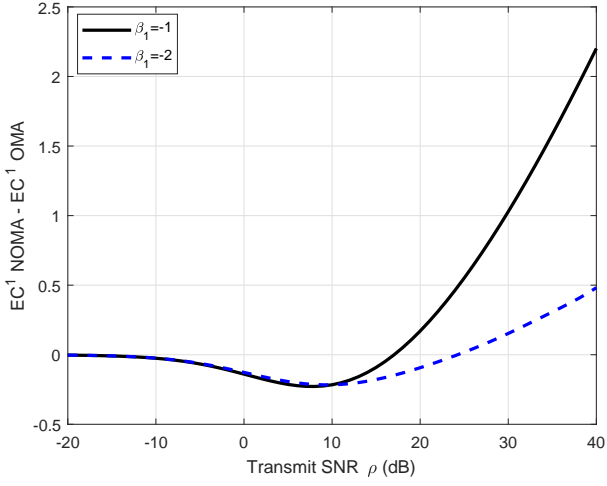}
      \caption{ $E_c^1-\Tilde{E}_c^1$ versus $\rho$, for several values of the normalized delay exponent.}
      \label{fig11}
     \end{center}
  \end{figure}

Fig. \ref{fig11} shows the difference of the EC in NOMA and the EC in OMA of the weak user. This curve starts initially at zero, then decreases to a certain minimum and starts increasing at the high values of transmit SNR. This confirms Lemma 2. When the delay is equal to $-1$, we see that for $\rho \in [-20,15]$ dB, the difference values are negative, indicating that OMA outperforms NOMA in this range. But when $\rho>15$ dB, the values are positive, i.e., NOMA offers better link-layer rates. However, the particular ranges depend not only on the delay exponents but also on the power allocation coefficients. By increasing the transmission power of the weak user and reducing the transmission power of the strong user, we notice that the range is reduced. That range expands when we do the inverse. Also, when the delay becomes more stringent, e.g., $\beta_1$=$\beta_2$=-2, the zero crossing moves from $15$ to $25$ dB.

\begin{figure}[t]
  \begin{center}
      \includegraphics[width=0.5\textwidth]{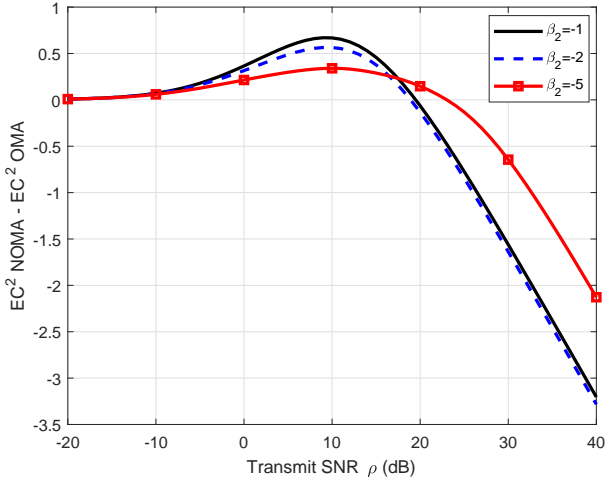}
      \caption{ $E_c^2-\Tilde{E}_c^2$ versus $\rho$, for various normalized delay exponent.}
      \label{fig12}
   \end{center}
  \end{figure}

Figure \ref{fig12} shows the difference of the EC in NOMA and the EC in OMA for the strong user. This curve starts initially at zero, then increases to a certain maximum and starts decreasing without bound at high values of the transmit SNR. This confirms Lemma 3. We note that the maximum of these curves decreases when the delay becomes more stringent. 

\begin{figure}[t]
   \begin{center}
      \includegraphics[width=0.5\textwidth]{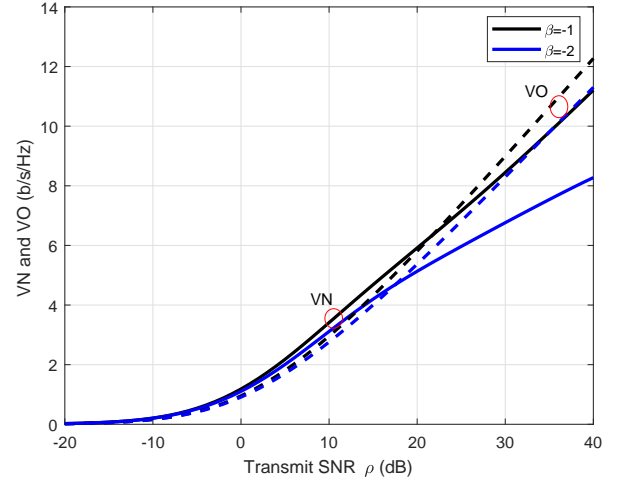}
      \caption{ $V_N$ and $V_O$ versus $\rho$, for several values of normalized delay exponent.}
      \label{fig13}
    \end{center}
  \end{figure}

 To investigate the impact of $\rho$ on the performance of the total link-layer rate for the two-user system, in Fig. \ref{fig13} the plots for $V_N$ in NOMA and $V_O$ in OMA, versus the transmit SNR are depicted for various delay exponents. The curves demonstrate that for both NOMA and OMA, the total EC for the two users starts at the initial value of 0 and then increases with the transmit SNR, as outlined in Lemma 4. When $\rho$ is very small, the total link-layer rate for the two user in NOMA, $V_N$, increases faster than $V_O$ in OMA. On the contrary, with the increase of the transmit SNR, $V_O$ becomes gradually higher than $V_N$.  At very high values of the transmit SNR, the gap between the sum EC with NOMA and OMA increases further. Finally, when the delay becomes more stringent, the sum EC of both NOMA and OMA decreases.
 
\begin{figure}[t]
     \begin{center}
      \includegraphics[width=0.5 \textwidth]{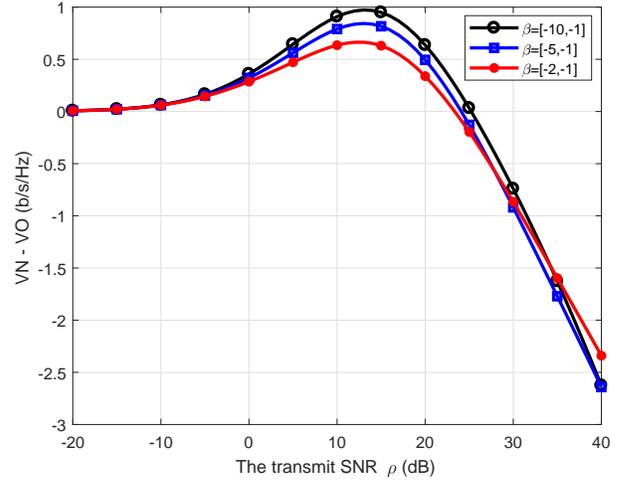}
      \caption{ $V_N$ - $V_O$ versus $\rho$ for various normalized delay.}
      \label{fig14}
      \end{center}
  \end{figure}

\begin{figure}[t]
   \begin{center}
      \includegraphics[width=0.5\textwidth]{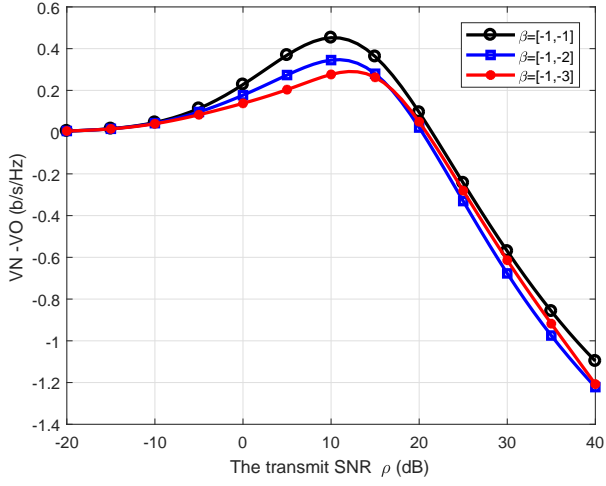}
      \caption{ $V_N$ - $V_O$ versus $\rho$ for various normalized delay.}
      \label{fig15}
     \end{center}
  \end{figure}

 Finally,  Figs \ref{fig14} and \ref{fig15} depict the sum ECs versus $\rho$, for several values of the (negative) normalized delay exponent. In Fig. \ref{fig14}, the delay of the strong user is fixed, while the delay exponent of the weak user varies. It is shown that in this case, the highest delay QoS (i.e., the smallest negative normalized delay exponent) of the weak user corresponds to the highest gap between the sum ECs $V_N-V_O$.  On the other hand, when the delay of the weak user is fixed, Fig. \ref{fig15} shows that the smallest delay Qos (i.e., the highest negative normalized delay exponent) for the strong user corresponds to the largest gap in $V_N-V_O$.  
  
  The curve of $V_N-V_O$ starts at zero, increases to a maximum, and returns to negative values. The transition to zero is at around $\rho$ = 25, and $\rho$ =20 respectively for the figures \ref{fig14} and \ref{fig15}. That means from 0 to 25dB (20dB in the Figure \ref{fig15}), the total link-layer rate of NOMA is higher than the OMA one. And when $\rho$ becomes larger than this transition point, the total link-layer rate of OMA outperforms the NOMA one.

\section{Conclusions and Future Work}
The concept of the EC enabled us to study the achievable data-link layer rates when the delay QoS guarantees are in place in the form of delay exponents. We  investigated the EC for the uplink of a two-user NOMA network, assuming a Rayleigh block fading channel. We derived novel closed-form expressions for the ECs of the two users and provided a comparison between NOMA and OMA.
In NOMA networks, we showed that the ECs of both users decrease as the delay constraints become stringent. On the other hand, at high transmit SNRs, the EC of the weak user can surpass the EC of the strong user, as the latter is  limited due to interference. This provides the possibility of switching between NOMA and OMA according to the individual users' delay constraints and transmit powers. In future work, the impact of user pairing will also be investigated. 

%\appendices
\section*{Appendix I}
For the second user, we have:
\begin{eqnarray}
  E_c^2&=&\frac{1}{\beta_2}\log_2\Bigg(2\int_{0}^{\infty}(\frac{\rho P_2}{1+\rho P_1 x_1})^{\beta_2}e^{-x_1} \nonumber\\ 
  & \times&\int_{x_{1}}^{\infty}\Big(\frac{1+\rho P_1 x_1}{\rho P_2}+ x_2\Big)^{\beta_2}e^{-x_2}dx_2 dx_1 \Bigg) \nonumber.
  %=\frac{1}{\beta_2}\log_2(\mathbf{E}[(1+ \frac{\rho P_2 x_{2}}{1+ \rho %P_1 x_{1}})^{\beta_2}])\\
  %&=\frac{1}{\beta_2}\log_2\left(\int_{0}^{\infty}\int_{x_{1}}^{\infty}\Big(1+ \frac{\rho P_2 x_2}{1+\rho P_1 x_1}\Big)^{\beta_2}f_{\gamma_{1:2}}(x_1,x_2)dx_2 dx_1\right)\\
  %=\frac{1}{\beta_2}\log_2\left(2\int_{0}^{\infty}\int_{x_{1}}^{\infty}\Big(1%+ \frac{\rho P_2 x_2}{1+\rho P_1 x_1}\Big)^{\beta_2}e^{-x_1}e^{-x_2}dx_2 %dx_1\right)\\
\end{eqnarray}
Set $z=\frac{1+\rho P_1 x_1}{\rho P_2}+ x_2$, which means we have $x_2=z-\frac{1+\rho P_1 x_1}{\rho P_2}$ and $dx_2=dz$. Then, \newline
%when $x_2\rightarrow x_1$, $z \rightarrow \frac{1+\rho P_1 x_1}{\rho P_2}+ x_1=\frac{1+\rho x_1}{\rho P_2}$ \newline
%When $x_2 \rightarrow \infty$, $z \rightarrow \infty$
\begin{eqnarray}
  E_c^2 %=\frac{1}{\beta_2}\log_2\left(2\int_{0}^{\infty}(\frac{\rho P_2}{1+\rho P_1 x_1})^{\beta_2}e^{-x_1}\int_{\frac{1+\rho x_1}{\rho P_2}}^{\infty}z^{\beta_2}\\e^{-(z-\frac{1+\rho P_1 x_1}{\rho P_2})}dz dx_1 \Big)\\
  \!\!\!\!  &=& \!\!\!\! \frac{1}{\beta_2}\log_2\Big(2e^{\frac{1}{\rho P_2}}\int_{0}^{\infty}(\frac{\rho P_2}{1+\rho P_1 x_1})^{\beta_2}e^{-x_1}e^{\frac{P_1 x_1}{ P_2}} \nonumber\\
  && \!\!\!\! \int_{\frac{1+\rho x_1}{\rho P_2}}^{\infty}z^{\beta_2}e^{-z}dz dx_1 \Big)\nonumber\\
  &=& \!\!\!\! \frac{1}{\beta_2}\log_2\Bigg(2(\rho P_2)^{\frac{\beta_2}{2}} e^{\frac{1}{2 \rho P_2}}\int_{0}^{\infty}(1+\rho P_1 x_1)^{-\beta_2} 
  \nonumber\\
  && \!\!\!\! \!\!\!\!   (1+\rho x_1)^{\frac{\beta_2}{2}}e^{\frac{(2P_1-2 P_2-1) x_1}{2 P_2}}\Big[\mathbf{W}_{\frac{\beta_2}{2},\frac{1+\beta_2}{2}}(\frac{1+\rho x_1}{\rho P_2}) \Big] dx_1 \Bigg)
  \nonumber \\
   &=& \!\!\!\! \frac{1}{\beta_2}\log_2\Bigg(2 P_2 (\rho P_2)^{\beta_2} e^{\frac{1}{\rho P_2}}e^{-\frac{(P_1-P_2)}{\rho P_2}}\nonumber\\
   && \!\!\!\! \int_{\frac{1}{\rho P_2}}^{\infty}P_2^{-\beta_2}(1+\rho P_1 y)^{-\beta_2} e^{(P_1-P_2)y}\Gamma(1+\beta_2,y) dy \Bigg),
  \nonumber
\end{eqnarray}
%we note that:
%$ \int_{a}^{\infty}\frac{e^{-x}}{x^b}dx=a^{-\frac{b}{2}} e^{-\frac{a}{2}}\mathbf{W}_{-\frac{b}{2},-\frac{1-b}{2}}(a)$
%\newline
where $\mathbf{W}$ is the Whittaker W function.
Using the binomial expansion, we have %\newline
$(1+\rho P_1 y)^{-\beta_2}=\sum_{j=0}^{-\beta_2}\binom{-\beta_2}{j}(\rho P_1 y)^{j}$.
%\newline
%Then, using Taylor series expansion, we have %\newline
% $ \exp{(P_1-P_2)y}= \exp(-(P_2-P_1)y)=\sum_{k=0}^{\infty}\frac{(-1)^k (P_2-P_1)^k}{k!}y^k$, which converges.
and we get the expression given in (\ref{eq:EC2}).

\section*{Appendix II}
%Here the proof of Lemma 1:
By inserting $\rho\rightarrow0$ into \eqref{eq:EC1} and \eqref{eq:EC2}, we get 1) of Lemma 1, i.e.,
\begin{equation}
   \lim\limits_{\rho\rightarrow0}(E_c^1-\widetilde{E}_c^{1})=\frac{1}{\beta_1}\log_2(\frac{\mathbb{E}[(1+\rho P_1 |h_1|^2)^{\beta_2}]}{\mathbb{E}[(1+\rho |h_1|^2)^{\frac{\beta_2}{2}}]})=0 \nonumber,
\end{equation}
\begin{equation}
   \lim\limits_{\rho\rightarrow0}(E_c^2-\widetilde{E}_c^{2})=\frac{1}{\beta_2}\log_2\left(\frac{\mathbb{E}[(1+\frac{\rho P_2 |h_2|^2}{1+\rho P_1 |h_1|^2})^{\beta_2}]}{\mathbb{E}[(1+\rho |h_2|^2)^{\frac{\beta_2}{2}}]}\right)=0 \nonumber.
\end{equation}

In the same way, by inserting $\rho\rightarrow \infty$ into \eqref{eq:EC1} and \eqref{eq:EC2}, we get 2) in Lemma 1, given below.

\begin{align}
\lim\limits_{\rho\rightarrow \infty}E_c^2&\rightarrow \frac{1}{\beta_2}\log_2(\mathbb{E}[(1+\frac{P_2 |h_2|^2}{P_1 |h_1|^2})^{\beta_2}]) \nonumber,\\
   \lim\limits_{\rho\rightarrow \infty}(E_c^1-\widetilde{E}_c^{1})&=\frac{1}{\beta_1}\log_2(\rho^{\frac{\beta_1}{2}} \frac{\mathbb{E}[(\frac{1}{\rho}+P_1 |h_1|^2)^{\beta_2}]}{\mathbb{E}[(\frac{1}{\rho}+ |h_1|^2)^{\frac{\beta_2}{2}}]})=\infty \nonumber,\\
   \lim\limits_{\rho\rightarrow \infty}(E_c^2-\widetilde{E}_c^{2})&=\frac{1}{\beta_2}\log_2(\frac{\mathbb{E}[(\frac{\frac{1}{\rho}+ P_1 |h_1|^2+ P_2 |h_2|^2}{\frac{1}{\rho}+ P_1 |h_1|^2})^{\beta_2}]}{\rho^{\frac{\beta_2}{2}}\mathbb{E}[(\frac{1}{\rho}+ |h_2|^2)^{\frac{\beta_2}{2}}]})= \!\!-\infty \nonumber.
   \end{align}

\section*{Appendix III}
To analyze the trends of $E_c^1$ and $\widetilde{E}_c^{1}$ with respect to $\rho$, we start with
%\begin{equations}
\begin{align}
    \frac{\partial E_c^1}{\partial\rho}&=\frac{1}{\beta_1 \ln2}\frac{\Big(\mathbb{E}[(1+\rho P_1 |h_1|^2)^{\beta_1}] \Big)'}{\mathbb{E}[(1+\rho P_1 |h_1|^2)^{\beta_1}]}\nonumber\\
    &=\frac{P_1}{\ln2} \frac{\mathbb{E}[|h_1|^2 (1+\rho P_1 |h_1|^2)^{\beta_1-1}]}{\mathbb{E}[(1+\rho P_1 |h_1|^2)^{\beta_1}]}\ge 0   \nonumber.
\end{align}
%\end{equations}
 %\newline 
Similarly, for user 1 in OMA we have
\begin{align}
    \frac{\partial \widetilde{E}_c^{1}}{\partial\rho}&=\frac{1}{\beta_1\ln2}\frac{\Big(\mathbb{E}[(1+\rho |h_1|^2)^{\frac{\beta_1}{2}}] \Big)'}{\mathbb{E}[(1+\rho |h_1|^2)^{\frac{\beta_1}{2}}]}\nonumber\\
    &=\frac{1}{2 \ln2} \frac{\mathbb{E}[|h_1|^2 (1+\rho |h_1|^2)^{\frac{\beta_1}{2}-1}]}{\mathbb{E}[(1+\rho |h_1|^2)^{\frac{\beta_1}{2}}]} \ge 0   \nonumber.
\end{align}
Then, we get that
\begin{align}
    \frac{\partial (E_c^1-\widetilde{E}_c^{1})}{\partial\rho}&=\frac{P_1}{\ln2} \frac{\mathbb{E}[|h_1|^2 (1+\rho P_1 |h_1|^2)^{\beta_1-1}]}{\mathbb{E}[(1+\rho P_1 |h_1|^2)^{\beta_1}]}\nonumber\\
    &-\frac{1}{2 \ln2} \frac{\mathbb{E}[|h_1|^2 (1+\rho |h_1|^2)^{\frac{\beta_1}{2}-1}]}{\mathbb{E}[(1+\rho |h_1|^2)^{\frac{\beta_1}{2}}]}\nonumber,
\end{align}
and
$\lim\limits_{\rho\rightarrow0}(\frac{\partial (E_c^1-\widetilde{E}_c^{1})}{\partial\rho})=\frac{(P_1-\frac{1}{2})}{\ln 2}\mathbb{E}[|h_1|^2] \leq0$. When $\rho>>1$, we have
\begin{align}
\frac{\partial (E_c^1-\widetilde{E}_c^{1})}{\partial\rho})&=\frac{P_1}{\ln2} \frac{\mathbb{E}[|h_1|^2 (\rho P_1 |h_1|^2)^{\beta_1-1}]}{\mathbb{E}[(\rho P_1 |h_1|^2)^{\beta_1}]}\nonumber\\
&-\frac{1}{2 \ln2} \frac{\mathbb{E}[|h_1|^2 (\rho |h_1|^2)^{\frac{\beta_1}{2}-1}]}{\mathbb{E}[(\rho |h_1|^2)^{\frac{\beta_1}{2}}]}=
\frac{1}{2\rho \ln2}\ge0 \nonumber.
\end{align}
When $\rho\rightarrow\infty$, this term approaches 0.

\section*{Appendix IV}
\begin{math}
E_c^2=\frac{1}{\beta_2}\log_2(\mathbb{E}[(1+ \frac{\rho P_2 |h_2|^2}{1+  \rho P_1 |h_1|^2})^{\beta_2}])\nonumber, \text{ and }
\end{math}
\begin{align}
    \frac{\partial E_c^2}{\partial\rho}&=\frac{1}{\beta_2 \ln2}\frac{\Big(\mathbb{E}[(1+ \frac{\rho P_2 |h_2|^2}{1+  \rho P_1 |h_1|^2})^{\beta_2}]\Big)'}{\mathbb{E}[(1+ \frac{\rho P_2 |h_2|^2}{1+  \rho P_1 |h_1|^2})^{\beta_2}]}\nonumber\\
     \!\!\!\!  \!\!\!\! &= \!\! \frac{1}{ \ln2}\frac{\mathbb{E}[\frac{P_2 |h_2|^2}{(1+\rho P_1 |h_1|^2)^2}(1+ \frac{\rho P_2 |h_2|^2}{1+  \rho P_1 |h_1|^2})^{\beta_2-1}]}{\mathbb{E}[(1+ \frac{\rho P_2 |h_2|^2}{1+  \rho P_1 |h_1|^2})^{\beta_2}]}\ge0 \nonumber.
\end{align} 
In the same way, for the user 2 in OMA, we have
\begin{align}
    \frac{\partial \widetilde{E}_c^{2}}{\partial\rho}&=\frac{1}{\beta_2\ln2}\frac{\Big(\mathbb{E}[(1+\rho |h_2|^2)^{\frac{\beta_2}{2}}] \Big)'}{\mathbb{E}[(1+\rho |h_2|^2)^{\frac{\beta_2}{2}}]}\nonumber\\
    &=\frac{1}{2 \ln2} \frac{\mathbb{E}[|h_2|^2 (1+\rho |h_2|^2)^{\frac{\beta_2}{2}-1}]}{\mathbb{E}[(1+\rho |h_2|^2)^{\frac{\beta_2}{2}}]} \ge 0   \nonumber, \text{ and}
\end{align}
\vspace{-0.2 cm}
\begin{align}
\frac{\partial (E_c^2-\widetilde{E}_c^{2})}{\partial\rho}&=\frac{1}{ \ln2}\frac{\mathbb{E}[\frac{P_2 |h_2|^2}{(1+\rho P_1 |h_1|^2)^2}(1+ \frac{\rho P_2 |h_2|^2}{1+  \rho P_1 |h_1|^2})^{\beta_2-1}]}{\mathbb{E}[(1+ \frac{\rho P_2 |h_2|^2}{1+  \rho P_1 |h_1|^2})^{\beta_2}]}\nonumber\\
&-\frac{1}{2 \ln2} \frac{\mathbb{E}[|h_2|^2 (1+\rho |h_2|^2)^{\frac{\beta_2}{2}-1}]}{\mathbb{E}[(1+\rho |h_2|^2)^{\frac{\beta_2}{2}}]} \nonumber.
\end{align}

When $\rho\rightarrow0$, we have that
$\lim\limits_{\rho\rightarrow0}(\frac{\partial (E_c^2-\widetilde{E}_c^{2})}{\partial\rho})=\frac{(P_2-\frac{1}{2})}{\ln 2}\mathbb{E}[|h_2|^2]$.
When $\rho$ is very large,
\begin{align}
&\frac{\partial (E_c^2-\widetilde{E}_c^{2})}{\partial\rho}=\frac{\mathbb{E}[\frac{P_2 |h_2|^2}{\rho^2 (\frac{1}{\rho}+ P_1 |h_1|^2)^2}(1+\frac{\rho}{\rho} \frac{(P_2 |h_2|^2)}{(\frac{1}{\rho}+ P_1 |h_1|^2)})^{\beta_2-1}]}{\ln2\mathbb{E}[(1+ \frac{\rho}{\rho}\frac{P_2 |h_2|^2}{(\frac{1}{\rho}+ P_1 |h_1|^2)})^{\beta_2}]}\nonumber\\ 
&\qquad\qquad -\frac{1}{2 \ln2}\frac{1}{\rho} \frac{\mathbb{E}[|h_2|^2 (\frac{1}{\rho}+ |h_2|^2)^{\frac{\beta_2}{2}-1}]}{\mathbb{E}[(\frac{1}{\rho}+ |h_2|^2)^{\frac{\beta_2}{2}}]}\nonumber\\
&=\!\!\frac{\mathbb{E}[\frac{P_2 |h_2|^2}{\rho^2 (P_1 |h_1|^2)^2}(1+\frac{P_2 |h_2|^2}{P_1 |h_1|^2})^{\beta_2-1}]}{\ln2\mathbb{E}[(1+\frac{P_2 |h_2|^2}{P_1 |h_1|^2})^{\beta_2}]} -\frac{1}{2 \ln2}\frac{1}{\rho} \frac{\mathbb{E}[( |h_2|^2)^{\frac{\beta_2}{2}}]}{\mathbb{E}[( |h_2|^2)^{\frac{\beta_2}{2}}]}\nonumber\\
&=\frac{P_2}{\rho^2 P_{1}^2 }\frac{\mathbb{E}[\frac{|h_2|^2}{(|h_1|^2)^2}(1+\frac{P_2 |h_2|^2}{P_1 |h_1|^2})^{\beta_2-1}]}{\ln2\mathbb{E}[(1+\frac{P_2 |h_2|^2}{P_1 |h_1|^2})^{\beta_2}]} -\frac{1}{2 \ln2}\frac{1}{\rho} \nonumber\\
&=\frac{\frac{P_2}{P_{1}^2\ln2} A-\frac{1}{2 \ln2}\rho}{\rho^2},\nonumber
\end{align}
where $A=\frac{\mathbb{E}[\frac{|h_2|^2}{(|h_1|^2)^2}(1+\frac{P_2 |h_2|^2}{P_1 |h_1|^2})^{\beta_2-1}]}{\mathbb{E}[(1+\frac{P_2 |h_2|^2}{P_1 |h_1|^2})^{\beta_2}]}$, unrelated to $\rho$. Hence, when $\rho$ is very large, $\frac{\partial (E_c^2-\widetilde{E}_c^{2})}{\partial\rho}$ can be approximated by $-\frac{1}{2 \ln2}\frac{1}{\rho}$, and it gradually approaches 0 when $\rho\rightarrow\infty$.

\section*{Appendix V}
%Here is provided the proof of the Lemma 4.\newline
Note that $V_N=E_c^1+E_c^{2}$. By using Lemma 1, we have 
$ \lim\limits_{\rho\rightarrow0}(V_N)=0$ and $ \lim\limits_{\rho\rightarrow\infty}(V_N)=\infty$. Then, we get that
\begin{align}
\frac{\partial V_N}{\partial\rho}&=\frac{\partial (E_c^1+E_c^{2})}{\partial\rho}=\frac{P_1}{\ln2} \frac{\mathbb{E}[|h_1|^2 (1+\rho P_1 |h_1|^2)^{\beta_1-1}]}{\mathbb{E}[(1+\rho P_1 |h_1|^2)^{\beta_1}]}\nonumber\\
&+\frac{1}{ \ln2}\frac{\mathbb{E}[\frac{P_2 |h_2|^2}{(1+\rho P_1 |h_1|^2)^2}(1+ \frac{\rho P_2 |h_2|^2}{1+  \rho P_1 |h_1|^2})^{\beta_2-1}]}{\mathbb{E}[(1+ \frac{\rho P_2 |h_2|^2}{1+  \rho P_1 |h_1|^2})^{\beta_2}]}\geq 0.\nonumber
\end{align}
%Which is non negative, because $ \frac{\partial E_c^1}{\partial\rho}\ge0$ and $ \frac{\partial E_c^2}{\partial\rho}\ge0$ as we previously showed.
%\newline

When $\rho\rightarrow 0$, we have
$ \lim\limits_{\rho\rightarrow 0}(\frac{\partial V_N}{\partial\rho})=\frac{P_1}{\ln2}\mathbb{E}[|h_1|^2]+\frac{P_2}{ \ln2}\mathbb{E}[|h_2|^2]$. 
%\newline
When $\rho\rightarrow\infty$, we get that
\begin{equation}
\lim\limits_{\rho\rightarrow\infty}\frac{\partial V_N}{\partial\rho}=\frac{1}{\rho \ln2}+\frac{\mathbb{E}[\frac{P_2 |h_2|^2}{(P_1 |h_1|^2)^2}(1+ \frac{P_2 |h_2|^2}{P_1 |h_1|^2})^{\beta_2-1}]}{\rho^2 \ln2\mathbb{E}[(1+ \frac{P_2 |h_2|^2}{ P_1 |h_1|^2})^{\beta_2}]} \nonumber  \!\!=\!\!0.
\end{equation} 
%Which equals to $0$.
%newline

For $V_O$ in the case of OMA, we note that
%\newline
$V_O=\widetilde{E}_c^{1}+\widetilde{E}_c^{2}$. By using Lemma 1, we have 
$ \lim\limits_{\rho\rightarrow0}(V_0)=0$ and $ \lim\limits_{\rho\rightarrow\infty}(V_0)=\infty$.
Then,
\begin{align}
\frac{\partial V_0}{\partial\rho}&=\frac{\partial (\widetilde{E}_c^{1}+\widetilde{E}_c^{2})}{\partial\rho}=\frac{1}{2 \ln2} \frac{\mathbb{E}[|h_1|^2 (1+\rho |h_1|^2)^{\frac{\beta_1}{2}-1}]}{\mathbb{E}[(1+\rho |h_1|^2)^{\frac{\beta_1}{2}}]}\nonumber \\
&+\frac{1}{2 \ln2} \frac{\mathbb{E}[|h_2|^2 (1+\rho |h_2|^2)^{\frac{\beta_2}{2}-1}]}{\mathbb{E}[(1+\rho |h_2|^2)^{\frac{\beta_2}{2}}]}\nonumber \geq 0.
\end{align}
%Which is non negative, because $ \frac{\partial\widetilde{E}_c^{1}}{\partial\rho}\ge0$ and $ \frac{\partial \widetilde{E}_c^{2}}{\partial\rho}\ge0$ as we previously showed.
%\newline
When $\rho\rightarrow0$, we have $ \lim\limits_{\rho\rightarrow 0}(\frac{\partial V_O}{\partial\rho})=\frac{1}{2 \ln2}\mathbb{E}[|h_1|^2]+\frac{1}{2  \ln2}\mathbb{E}[|h_2|^2]$. 
%\newline
When $\rho\rightarrow\infty$, we have that $\lim\limits_{\rho\rightarrow\infty}(\frac{\partial V_O}{\partial\rho})= \lim\limits_{\rho\rightarrow\infty}(\frac{1}{2 \rho \ln2}+\frac{1}{2 \rho \ln2})=\lim\limits_{\rho\rightarrow\infty}(\frac{1}{ \rho \ln2})$,
which equals to $0$. %\newline

\bibliographystyle{IEEEtran}
\bibliography{IEEEabrv,biblio}
\balance
%\onecolumn
%\input{Appendix}

\end{document}